\documentclass[twocolumn,aps,prb,groupedaddress,superscriptaddress,amsmath,amssymb,showpacs,noeprint,english]{revtex4-2}

\usepackage{graphicx}
\usepackage{dcolumn}
\usepackage{bm}
\usepackage[T1]{fontenc}
\usepackage{mathrsfs}
\usepackage{amsbsy}
\usepackage{amstext}
\usepackage{amssymb}
\usepackage{setspace}
\usepackage{esint}
\usepackage{xcolor,colortbl}
\usepackage{float}
\usepackage[colorlinks,citecolor=blue,urlcolor=blue,linkcolor=blue,hypertexnames=true]{hyperref} 
\usepackage{babel}
\usepackage{booktabs}
\usepackage{tikz}

\usetikzlibrary{svg.path}

\definecolor{orcidlogocol}{HTML}{A6CE39}
\tikzset{
  orcidlogo/.pic={
    \fill[orcidlogocol] svg{M256,128c0,70.7-57.3,128-128,128C57.3,256,0,198.7,0,128C0,57.3,57.3,0,128,0C198.7,0,256,57.3,256,128z};
    \fill[white] svg{M86.3,186.2H70.9V79.1h15.4v48.4V186.2z}
                 svg{M108.9,79.1h41.6c39.6,0,57,28.3,57,53.6c0,27.5-21.5,53.6-56.8,53.6h-41.8V79.1z M124.3,172.4h24.5c34.9,0,42.9-26.5,42.9-39.7c0-21.5-13.7-39.7-43.7-39.7h-23.7V172.4z}
                 svg{M88.7,56.8c0,5.5-4.5,10.1-10.1,10.1c-5.6,0-10.1-4.6-10.1-10.1c0-5.6,4.5-10.1,10.1-10.1C84.2,46.7,88.7,51.3,88.7,56.8z};
  }
}

\newcommand\orcidicon[1]{\href{https://orcid.org/#1}{\mbox{\scalerel*{
\begin{tikzpicture}[yscale=-1,transform shape]
\pic{orcidlogo};
\end{tikzpicture}
}{|}}}}

\begin{document}

\title{Quantum Phases Classification Using Quantum Machine Learning\\ with SHAP-Driven Feature Selection}

\author{Giovanni S. Franco}
\affiliation{S{\~a}o Paulo State University (UNESP), School of Sciences, 17033-360 Bauru, S{\~a}o Paulo, Brazil}

\author{Felipe Mahlow}
\affiliation{S{\~a}o Paulo State University (UNESP), School of Sciences, 17033-360 Bauru, S{\~a}o Paulo, Brazil}
	
\author{Pedro M. Prado}
\affiliation{S{\~a}o Paulo State University (UNESP), School of Sciences, 17033-360 Bauru, S{\~a}o Paulo, Brazil}

\author{Guilherme E. L. Pexe}
\affiliation{S{\~a}o Paulo State University (UNESP), School of Sciences, 17033-360 Bauru, S{\~a}o Paulo, Brazil}

\author{Lucas A. M. Rattighieri}
\affiliation{S{\~a}o Paulo State University (UNESP), School of Sciences, 17033-360 Bauru, S{\~a}o Paulo, Brazil}

\author{Felipe F. Fanchini}
\email{felipe.fanchini@unesp.br}
\affiliation{S{\~a}o Paulo State University (UNESP), School of Sciences, 17033-360 Bauru, S{\~a}o Paulo, Brazil}
\affiliation{QuaTI - Quantum Technology \& Information, 13560-161 São Carlos-SP, Brazil}

\date{\today}

\begin{abstract}

In this study, we present an innovative methodology to classify quantum phases within the ANNNI (Axial Next-Nearest Neighbor Ising) model by combining Quantum Machine Learning (QML) techniques with the Shapley Additive Explanations (SHAP) algorithm for feature selection and interpretability. Our investigation focuses on two prominent QML algorithms: Quantum Support Vector Machines (QSVM) and Variational Quantum Classifiers (VQC). By leveraging SHAP, we systematically identify the most relevant features within the dataset, ensuring that only the most informative variables are utilized for training and testing. The results reveal that both QSVM and VQC exhibit exceptional predictive accuracy when limited to 5 or 6 key features, thereby enhancing performance and reducing computational overhead. This approach not only demonstrates the effectiveness of feature selection in improving classification outcomes but also offers insights into the interpretability of quantum classification tasks. The proposed framework exemplifies the potential of interdisciplinary solutions for addressing challenges in the classification of quantum systems, contributing to advancements in both machine learning and quantum physics.
\\

\end{abstract}

\maketitle
\section{Introduction}
Classifying quantum phases is a central problem in the fields of many-body physics and materials science \cite{Canabarro2019}, which has attracted renewed interest in recent years \cite{Paschen2021}.  In this context, low-dimensional quantum lattice models have emerged as essential theoretical frameworks, clarifying these complex phenomena \cite{Dagotto1994,Dzero2016}. As these theoretical approaches have evolved, machine learning (ML) techniques have gained prominence as a promising methodology for classifying and detecting quantum phase transitions \cite{Hastings2013,Schuch2011,Chen2011,Carleo2017,Huang2021}. Unlike traditional statistical models, ML algorithms have the ability to increase their predictive capabilities through data-driven learning \cite{Schuld2018, Mehta2019}. 
These techniques are now applied in various scientific domains, offering novel methods for identifying and analyzing quantum phase transitions \cite{Carrasquilla2017, Dong2019, Shiina2020, Rem2019, Broecker2017, Canabarro2019, Polet2020, Yu2021, Mahlow_2023}.

With the advancement of quantum technologies, especially in quantum computing, the scientific and technological community is actively exploring the intersection of machine learning and quantum computing, a field known as QML \cite{Schuld_book_2018}. QML presents significant promise for the future, with potential applications in the natural sciences \cite{PERALGARCIA2024100619}. It takes advantage of the high expressivity of parameterized quantum circuits as statistical models \cite{Monaco_2023}. Despite its potential advantages in specific applications \cite{Huang2021b, Glick_2024}, QML faces significant challenges in the Noisy Intermediate-Scale Quantum (NISQ) era \cite{preskill2018quantum}, where the availability of practical quantum computers remains very limited. In this sense, the precise advantages of quantum computers in the era of deep neural networks remain undefined \cite{Monaco_2023}, making the future of QML both promising and uncertain.


One of the primary challenges for practical QML is the number of qubits required. Indeed, large datasets and complex learning processes involving numerous features demand a substantial number of qubits. In this sense, this highlights the importance of studying the relevance of the features, as identifying the most influential features can optimize the performance of QML models and enhance their practical applications. In this context, Explainable Artificial Intelligence (XAI) algorithms \cite{e23010018} can contribute to addressing some of the challenges faced by QML. 
Among XAI methods, Shapley Additive Explanations (SHAP) is a particularly effective tool \cite{NIPS2017_8a20a862}. SHAP quantifies each feature's contribution to model predictions, aiding in identifying influential features. 
This capability is especially valuable in the NISQ era, where optimizing feature selection can help overcome computational limitations and improve the practical applicability of QML algorithms.

In this study, we examine the axial next-nearest neighbor Ising (ANNNI) model \cite{SELKE1988213, chakrabarti2008quantum}. This model is particularly significant because of its role as a simple system that integrates quantum fluctuations with frustrated competing exchange interactions \cite{Canabarro2019}. We propose a method to accurately classify quantum phases within the ANNNI model using SHAP with Quantum Support Vector Machines and Variational Quantum Classifiers. Using the SHAP algorithm, we identify the most influential features, and through detailed data preprocessing and feature mapping, we prepare our dataset for effective classification.  The application of QSVM and VQC allows us to demonstrate the potential of these QML models to classify the phases in the ANNNI model effectively. We also studied the impact of the number of features on the precision of both algorithms \cite{havlivcek2019supervised}.  This approach not only highlights the potential of QML models to analyze complex quantum systems but also examines the impact of feature selection on the precision of these models.

This paper is organized as follows. Section \ref{annni} introduces the ANNNI model, emphasizing its importance and detailing its phases. Section \ref{shap} presents the SHAP method. Section \ref{c.qml} examines the QML methods employed, specifically QSVM and VQC. Section \ref{methods} details the methodology for feature selection, data preprocessing, feature mapping, and classification tasks using the QML algorithms. Section \ref{results} presents the results obtained from the computational simulations, and Section \ref{conclusion} summarizes our findings, highlighting potential future research and the implications of our work in quantum phase transitions.

\section{The ANNNI Model} 
\label{annni}

The spin-1/2 {Axial (or anisotropic) Next-Nearest Neighbor Ising} model, commonly referred to as ANNNI, was employed in this study. The phase transitions in the ANNNI model are governed by the following Hamiltonian:
\begin{equation}
H=-J\sum_{j=1}^{N}(\sigma_j^z\sigma_{j+1}^z - \kappa\sigma_j^z\sigma_{j+2}^z+g\sigma_{j}^x).
\end{equation}
In the above expression, the Pauli matrices $\sigma^a_j$, with $a=x,z$, act on the spin-1/2 degree of freedom of the $j$-th site. The coupling constant $J>0$, which sets the energy scale (we use $J=1$), characterizes the nearest-neighbor ferromagnetic exchange interaction, $\kappa$ is the dimensionless coupling constant for the next-nearest neighbor interaction, and $g$ represents the transverse magnetic field. The ground-state phase diagram of the ANNNI model includes ferromagnetic, antiphase, paramagnetic, and floating phases \cite{Canabarro2019, yang2023pattern}. This diagram results from the frustration between the nearest-neighbor ferromagnetic interactions and the next-nearest-neighbor antiferromagnetic interactions, compounded by the disordering effects introduced by the transverse field.

The ferromagnetic phase occurs when the nearest-neighbor interaction is predominant, typically resulting in states where spins align in a single direction ($\uparrow \uparrow \uparrow \uparrow \uparrow \uparrow \uparrow \uparrow$).  When the influence of the next-nearest-neighbor interaction prevails, it leads to the antiphase, characterized by a periodicity of four sites ($\uparrow \uparrow \downarrow \downarrow \uparrow \uparrow \downarrow \downarrow$). In the presence of a strong transverse field, the spins tend to align with the field, resulting in the paramagnetic phase. The floating phase \cite{Rieger_1996} is characterized by algebraically decaying spin correlations, indicating a modulated long-range order that varies with the frustration parameter $\kappa$, despite lacking a stable long-range order.

To generate the training data for our QML models, we calculated the pairwise correlations of the ground state, the lowest energy state of the Hamiltonian. Specifically, since we are considering a closed chain (where the first and last spins are connected, forming a loop), we computed the expected values of the observables $(\langle\sigma_{i}^{x} \sigma_{j}^{x}\rangle, \langle\sigma_{i}^{y} \sigma_{j}^{y}\rangle, \langle\sigma_{i}^{z} \sigma_{j}^{z}\rangle)$ for $i=1$ and $j = [2, N/2+1]$, where $N$ is the number of spins. Note that calculating only up to $j = N/2+1$ for $i=1$ avoids redundancy, as those would duplicate the obtained data. Thus, for systems with 8 and 12 sites, we obtain 12 and 18 features, respectively. By incrementing the values of $g$ and $\kappa$ in steps of $0.01$ across the range from $[0, 1]$, we calculated $10.000$ data points. These data points are subsequently analyzed using machine learning techniques.

Given that supervised learning models are employed for data classification within the ANNNI model framework, a thorough examination of phase transitions is important for our analysis. These transitions clearly delineate the distinct phases the system undergoes under different parameter regimes. In this sense, a comprehensive understanding of these boundaries is important for precise identification and categorization of phases, an essential ingredient for effectively training the models. In the limiting case where $\kappa = 0$, the model reduces to the well-known transverse field Ising model, which allows for exact analytical solutions \cite{Sachdev2011}. At the critical point $g = 1$, the system undergoes a second-order phase transition, delineating a shift from a ferromagnetic phase ($g < 1$) to a paramagnetic phase ($g > 1$). 
An approximate analytical expression for the critical value of $g$ within the parameter range $0 \leq \kappa \leq 1/2$ is provided by \cite{suzuki2012quantum}: 
\begin{equation} 
g_I(\kappa) \approx \frac{1 - \kappa}{\kappa} \left( 1 - \sqrt{\frac{1 - 3\kappa + 4\kappa^2}{1 - \kappa}} \right) 
\end{equation}

For $\kappa > 1/2$, the model undergoes a Berezinsky-Kosterlitz-Thouless transition from the floating phase to the paramagnetic phase as $g$ increases. The approximate critical value of $g$ for this transition is given by \cite{PhysRevB.76.094410}:
\begin{equation}
g_{BKT}(\kappa) \approx 1.05 \sqrt{(\kappa - 0.5)(\kappa - 0.1)}
\end{equation}
As discussed, these transitions are fundamental to our supervised learning approach, as they delineate the boundaries between different phases of the model. Accurate classification of the data by our QML algorithms depends on a precise understanding of these phase boundaries.

\section{Shapley Additive Explanations}
\label{shap}

Shapley Additive Explanations is an approach in the field of Explainable AI that provides a unified framework for interpreting the output of machine learning models \cite{NIPS2017_8a20a862}. It is based on cooperative game theory and assigns each feature in a prediction an importance value representing its contribution to the model's output \cite{10.1007/978-3-030-57321-8_2}. The core idea behind SHAP is to decompose the model's prediction into contributions from individual features, allowing for a more granular understanding of how each feature influences the final outcome. The significance of SHAP lies in its ability to provide both local and global explanations for model predictions. Local explanations focus on explaining the prediction of a single instance by attributing the contribution of each feature to that specific prediction. Global explanations, on the other hand, provide an overview of feature importance across the entire dataset, allowing users to understand which features have the most significant impact on the model's overall behavior \cite{lundberg2020local}.

In practical terms, SHAP produces visualizations such as SHAP summary plots, SHAP dependence plots, and individual SHAP value plots to help users interpret model predictions \cite{lundberg2018consistent}. These visualizations provide insights into how each feature affects the model's output and can aid in debugging models, understanding model behavior, and gaining trust in AI systems. Despite its effectiveness, SHAP is not without limitations. It can be computationally expensive, especially for large datasets and complex models, and SHAP do not address the problem of fairness directly \cite{10.1007/978-3-030-67664-3_11}. Additionally, interpreting the output of SHAP requires some understanding of the underlying machine learning model and may not always provide intuitive explanations for non-technical users.

In this work, we employed SHAP to determine feature importance in our classification tasks, an essential step as feature selection reduces the number of qubits required for QML, which is particularly relevant in the current NISQ era. Although recent studies have begun exploring the intersection of explainable AI and QML  \cite{power2024featureimportanceexplainabilityquantum, steinmüller2022explainableaiquantummachine}, there remains a lack of a consolidated framework in this domain. Consequently, to ascertain the importance of features when classifying quantum phases, we initially computed the feature importance through SHAP by applying a classical algorithm, the Support Vector Classifier (SVC), and we use it's selected most important features in order to train the quantum algorithms (QSVM and VQC). It is important to emphasize that, training SHAP directly on a quantum algorithm is, at present, extremely costly in terms of computational time, making feature selection using a classical algorithm the only practical approach for identifying relevant features. Nonetheless, as demonstrated in our results, this methodology already yields promising outcomes


\section{Quantum Machine Learning} 
\label{c.qml}

Quantum Machine Learning is an emerging field that combines quantum computing principles with machine learning techniques. By harnessing the unique properties of quantum computing, such as superposition, entanglement, and quantum parallelism, QML algorithms aim to solve complex problems more efficiently than their classical counterparts, such as data analysis and pattern recognition \cite{weinberg2024quantumalgorithmsnewfrontier}. This section delves into this interplay, exploring techniques such as variational quantum circuits, quantum kernels and quantum support vector machines, which utilize quantum circuits for kernel computations and optimization.

\subsection{Quantum Support Vector Machine}
\label{qsvm}

The Support Vector Machine (SVM) is a supervised learning algorithm used for classification and regression tasks \cite{708428}. In classification, the SVM works by finding the hyperplane that best separates different classes in the feature space, maximizing the margin between them. This involves seeking the largest possible distance between the separating hyperplane and the nearest data points of each class, known as support vectors \cite{cortes1995support, burges1998tutorial}. This approach improves the model’s ability to generalize effectively to new, unseen data.

When combined with kernel methods, the SVM can effectively handle data that is not linearly separable. Kernel methods map the input data to a higher-dimensional space where linear separation becomes feasible. The most commonly used kernel is the radial basis function (RBF) kernel, defined as: \begin{equation}
K(\mathbf{x}_i, \mathbf{x}_j) = \exp(-\gamma \|\mathbf{x}_i - \mathbf{x}_j\|^2)
\end{equation}
where $\gamma$ is a kernel parameter that controls the influence of each data point \cite{burges1998tutorial}.  This technique allows the SVM to create complex decision boundaries that are suitable for high-dimensional data and nonlinear problems. By using kernels, the original problem of linear separation in the input space is transformed into a problem of linear separation in the higher-dimensional feature space, allowing for nonlinear separations in the original space.

Kernel methods employ functions known as \textit{kernels} to measure distances in the input space, essential for data classification by mapping points to a higher-dimensional space. The \textit{representer theorem} and the \textit{kernel trick} allow many models to be expressed and constructed in terms of these kernels \cite{schuld2021supervised}. A kernel $\kappa : X \times X \rightarrow \mathbb{C}$ is a positive semidefinite function used to compute inner products of data within the space $X$, a fundamental concept in kernel methods, which can be extended to a Reproducing Kernel Hilbert Space \cite{Schuld_book_2018}.

Quantum kernels, on the other hand, map inputs to vectors in a Hilbert space, using quantum states $|\phi(x)\rangle$ as \textit{feature maps}. The inner product between these states defines the quantum kernel: 
\begin{equation}
\kappa(x, x') = \langle \phi(x) | \phi(x') \rangle.\label{qk}
\end{equation}
Quantum devices can compute this inner product, which can then be applied in classical machine learning algorithms. Identifying useful and advantageous quantum kernels remains a significant challenge, with the input encoding strategy largely determining the nature of the quantum kernel \cite{Schuld_book_2018}. The QSVM extends classical SVM methods by leveraging the quantum \textit{kernel trick}, where data is mapped to a higher-dimensional Hilbert space through quantum states, enabling more efficient kernel calculations. 

\subsection{Variational Quantum Classifier}
The Variational Quantum Classifier is a quantum algorithm that utilizes parameterized quantum circuits and classical optimization to perform classification tasks. These algorithms are used to categorize data into distinct classes and are especially well-suited for NISQ devices.
As a type of variational quantum algorithm (VQA), it employs parameterized quantum circuits, represented by vectors \(\boldsymbol{\theta}\), which are associated with a cost function \(C(\boldsymbol{\theta})\) that must be minimized. The objective is to find the optimal set of parameters \(\boldsymbol{\theta}^*\) that minimizes \(C(\boldsymbol{\theta})\) \cite{Cerezo_2021}.


{An essential component of VQAs is the selections of an ansatz architecture, which defines the initial quantum circuit with adjustable parameters. Standard ansatz structures include the Hardware Efficient Ansatz \cite{Kandala_2017}, Unitary Coupled Cluster Ansatz \cite{Romero_2018}, Quantum Alternating Operator Ansatz \cite{Hadfield_2019}, and Variational Hamiltonian Ansatz \cite{Wecker_2015}. Additionally, VQAs depend on classical optimizers to fine-tune the parameters of the quantum circuits. Techniques such as Gradient Descent \cite{Ruder_2016}, Adam \cite{Kingma_2014}, Simultaneous Perturbation Stochastic Approximation (SPSA) \cite{Spall_1992}, and Meta-Learning \cite{Finn_2017} are frequently used for this task \cite{Cerezo_2021}.}
By optimizing the circuit parameters, VQCs learn to distinguish between different categories, offering the potential to leverage quantum computing for solving complex classification problems more efficiently than classical approaches.

\section{Methods}
\label{methods}

In this section, we outline the methodology used in this study. We begin with feature selection using SHAP values to identify the most relevant features, followed by data preprocessing steps, including train-test splitting and feature scaling. Next, We describe the use of quantum feature maps, such as the ZFeatureMap \cite{IBM_2021}, to transform classical data into quantum states. Finally, we present the Quantum Support Vector Classifier (QSVC) and the Variational Quantum Classifier to predict the phase diagram of the ANNNI model.

\subsection{Feature Selection with SHAP}


In this work, we use the SHAP algorithm to identify the most relevant features in the learning process. This identification is based on a machine learning model, in this case the SVC, which assesses the importance of each feature in contributing to the predictions. SHAP quantifies how much each feature adds to or subtracts from the predicted value, providing a fair attribution of importance among features. In SHAP, this attribution is represented as an additive feature attribution model, expressed by the following linear equation \cite{NIPS2017_8a20a862}:
\begin{equation}
g(z') = \phi_0 + \sum_{j=1}^{M} \phi_j z'_j
\end{equation}
where $g(z')$ represents the explanation model, which approximates the output of the original model in an interpretable and transparent manner based on individual feature contributions. In this equation, $\phi_0$ is the model's baseline prediction, or the expected output when no feature values are present; in other words, $\phi_0$ is the mean value of all predictions across the dataset. The parameter $M$ denotes the total number of features in the model, and each $\phi_j$ represents the Shapley value for feature $j$, indicating its specific contribution to the prediction. Finally, $z'_j$ is a binary indicator for feature $j$, with $1$ indicating the feature's presence and $0$ indicating its absence.

The explanation model $g(z')$ is designed to provide a linear approximation of the learning model’s behavior for a specific input. Essentially, the idea is to replace the original feature vector $x$ with a binary string, $x'$, that represents the presence or absence of each feature. This simplified input $x'$ is then mapped back to the original input space using a function $h_x(x')$, which is specific to each input $x$. In fact, when $x'$ is a complete string of ones (indicating all features are present), the mapping ensures that $f(x) = f(h_x(x'))$, where $f(x)$ represents the predictive function of the original model. The explanation model $g(z')$ is constructed such that $g(z') \approx f(h_x(z'))$ whenever $z'\approx x'$, where $z'$ is a binary string that represents the subset of features selected for the approximation.

Many explainable methods obey the additive feature attribution model, in which the prediction is decomposed into linear contributions from each feature \cite{ribeiro2016why, shrikumar2017learning, bach2015pixel}, with the primary difference lying in how each feature's contribution, $\phi_j$, is computed. The true innovation of SHAP lies, in fact, in the rigorous and unified way it calculates the values $\phi_j$ for each feature. These values, called Shapley values, are calculated to ensure a fair and unique attribution of importance to each feature, adhering to three desirable properties: local accuracy, missingness, and consistency. Formally, Shapley values $\phi_j$ are calculated as the average of the marginal contributions of a feature across all possible feature combinations. Thus, given a set of features, the Shapley value of each reflects the average contribution of the feature to the outcome across different ``coalitions'' or subsets of features. This calculation allows a fair attribution of importance to each feature based on its average individual contribution. 

The classic Shapley value is defined based on cooperative game theory and is given by the equation \cite{NIPS2017_8a20a862}:
\begin{equation}
\phi_j = \sum_{S \subseteq N \setminus \{j\}} \frac{|S|!(|N| - |S| - 1)!}{|N|!} \left( v(S \cup \{j\}) - v(S) \right)\label{shapeq}
\end{equation}
where $S$ represents a subset of features without feature  $j$, $N$ is the total set of features, $v(S)$ is the value of the function (or model prediction) for the set $S$, and the combinatorial factors weight the contributions of each subset, ensuring that all combinations are considered. Despite the formal solution provided by Eq. (\ref{shapeq}), calculating the exact Shapley values by evaluating all possible combinations of features is computationally expensive. Thus, a practical approximation, introduced in \cite{NIPS2017_8a20a862}, is the Kernel SHAP method, which we use in our analysis. Kernel SHAP combines ideas from the LIME method \cite{ribeiro2016why} with classic Shapley values, providing an efficient approximation that avoids the need to evaluate all feature subset permutations. As we emphasize, like LIME, the SHAP model uses a locally weighted linear regression to approximate the model’s behavior around a specific instance. The key difference lies in the choice of sample weights and cost function, which are specifically adjusted to ensure that Kernel SHAP satisfies the three desirable properties of Shapley values. This method helps identify the influence of individual features on the model's predictions, allowing us to understand the relative importance of each feature. This information is visualized in a SHAP summary plot (see the Results section), which highlights each feature's contribution to the predictions and supports feature selection and optimization strategies. We focus on the features with the highest importance scores, aiming to improve the efficiency of the QML algorithms by reducing the feature set to the most relevant ones.



\subsection{Data Pre-processing}

As usual in supervised machine learning strategies, the dataset is divided into training and testing sets, with 30\% of the data allocated to the test set. To ensure that all features contribute equally to the model's performance and to improve convergence during training, feature scaling is applied to both the training and testing sets, adjusting them to a range between 0 and 1.
The scaling transformation is given by:
\begin{equation}
X' = \frac{X - X_{\min}}{X_{\max} - X_{\min}},
\end{equation}
where $X$ represents the original feature values, $X_{\min}$ and $X_{\max}$ are the minimum and maximum values of the features in the training set, respectively, and $X'$ represents the scaled feature values. 

\subsection{Feature Map}
In quantum machine learning, feature maps are critical for enhancing the representational capacity of quantum algorithms. They acts as intermediaries between classical data and quantum states, enabling the extraction of intricate features from the input data. A quantum feature map encodes classical data into a quantum state through a parameterized circuit, where data $x$ serve as parameters that are converted into quantum states $|\phi(x)\rangle$. Given their central role in QML, feature maps are an active area of research focused on developing new and improved techniques  \cite{9689853}. Typically, they are designed to produce specific entanglement structures and may incorporate repetitions to capture higher-order correlations.

In this study, we employ the {ZFeatureMap} \cite{IBM_2021} as the selected feature map for encoding classical data into quantum states in both the VQC and QSVM cases. The {ZFeatureMap} is a specialized instance of the \textit{Pauli Expansion Circuit}, facilitating the encoding of classical input data $\vec{x} \in \mathbb{R}^n$ into a quantum state. This feature map applies rotations based on the Pauli-Z operator.
The {ZFeatureMap} is particularly effective for encoding data through simple, localized operations. Moreover, it enables the representation of complex feature interactions through repeated applications of this transformation, enabling the detection of not only individual features but also the intricate ways in which different features combine and influence outcomes.  Repeated applications allow the quantum circuit to model higher-order correlations between features, enhancing its ability to capture complex, non-linear patterns in the data.

The feature map on $n$-qubits is generated by $\mathcal{U}_{\Phi(\vec{x})} = [U_{\Phi(\vec{x})} H^{\otimes n}]^p$, where $H$ is the Hadamard gate, $U_{\Phi(\vec{x})}$ applies the Pauli-based phase encoding to the $n$-qubit system, and $p$ represents the number of layers (in this study, $p=3$). Specifically,  
\begin{equation}
U_{\Phi(\vec{x})} = \exp\left(i\sum_{k=1}^n 2 x[k] \sigma^z_k \right),
\end{equation}  
where $x[k]$ corresponds to one of the feature values. 
While this feature map structure is effective, it is important to note that more complex ansatz structures can be designed to introduce entanglement, thereby increasing  the complexity of the feature map. In our case, we focus on single-qubit Pauli-$Z$ rotations, which provide an efficient method of data encoding without requiring multi-qubit interactions, as seen in more entangled maps such as the {ZZFeatureMap} \cite{qiskit2024}.

\subsection{Applying the QML Algorithms}

As previously mentioned, we employed two distinct quantum machine learning methods: the QSVM and the VQC. For the Quantum Support Vector Classification algorithm, a fidelity-based statevector kernel, Eq. (\ref{qk}), is used to construct the quantum kernel, which serves as the key component of the model. In this approach, only the kernel is quantum. Once the quantum kernel is calculated, the learning process proceeds as in classical SVM. As we can see, the kernel function is defined by the overlap of two simulated quantum statevectors generated by a parameterized quantum circuit, often referred to as the feature map. As usual, after training, predictions are made on the test dataset, and their accuracy is assessed by comparing them with the truth labels. The resulting accuracy is then carefully analyzed to evaluate the model's performance \cite{qiskit2024}.

The VQC, on the other hand, is a quantum machine learning algorithm that utilizes variational circuits to perform classification tasks. In our study, it employs the same quantum feature map used in the QSVM, which encodes classical data into quantum states, and is further combined with a parameterized quantum circuit (ansatz) that processes these states. During training, the parameters of the ansatz are adjusted using optimization techniques, such as the Simultaneous Perturbation Stochastic Approximation (SPSA) \cite{Spall_1992}, to minimize a cost function associated with classification accuracy. By learning optimal parameters, the VQC can effectively separate different classes in the quantum state space, leveraging the capability of quantum circuits to capture complex patterns and correlations within the data.



Following the application of the ZFeatureMap, the VQC employs the EfficientSU2 ansatz. This ansatz is designed to create entanglement among qubits through a sequence of single-qubit rotations and entangling gates. It efficiently explores the unitary group $SU(2^n)$, where $n$ is the number of qubits, allowing the circuit to represent a broad range of quantum states. The EfficientSU2 circuit applies layers of parameterized single-qubit rotations around the $Y$ and $Z$ axes, interleaved with entangling gates such as controlled-NOT (CNOT) gates. The general form of the ansatz can be expressed as:
\begin{equation}
U(\vec{\theta}) = \left( \prod_{l=1}^L U_{\texttt{ent}} U_{\texttt{rot}}(\vec{\theta}_{(l)}) \right),
\end{equation}
where $L$ is the number of layers, $U_{\texttt{rot}}(\vec{\theta}_{(l)})$ represents the parameterized single-qubit rotations in layer $l$, and $U_{\texttt{ent}}$ represents the entangling operations between qubits.

\section{Results} 
\label{results}
In this section, we first discuss the selection of the most important features using the SHAP algorithm. We then present the classification results for both the Quantum Support Vector Classification and the Quantum Variational Classifier models. Finally, we analyze the accuracy of both models as a function of the number of features selected through SHAP, demonstrating the effectiveness of this approach in improving the performance of quantum classifiers.

\subsection{Feature Selection}

\begin{figure}
    \centering
    \includegraphics[width=1\linewidth]{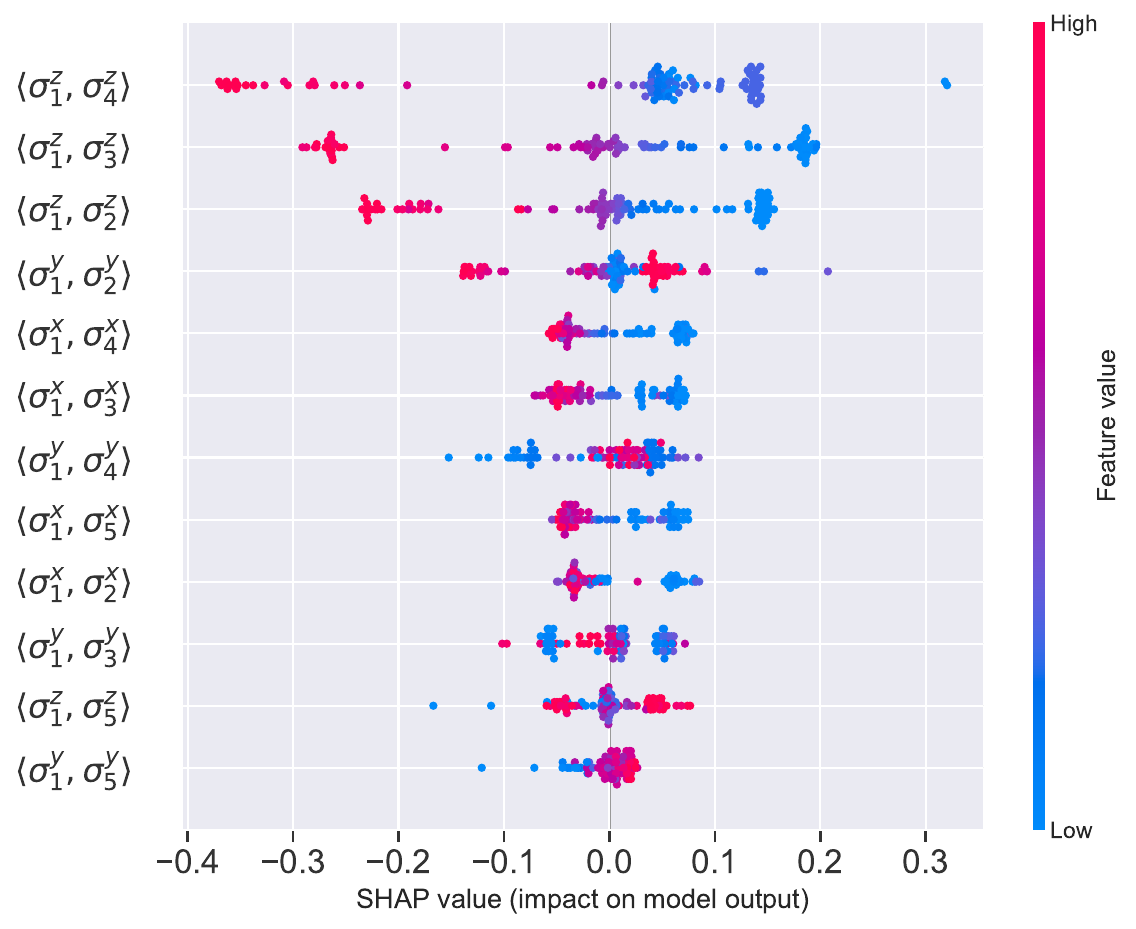}
    \caption{SHAP feature importance ranking for an 8-site ANNNI model using the Support Vector Classifier. The plot shows the relative importance of different features in classifying the phases of the model.}
    \label{fig:shap1}
\end{figure}

In our investigation, a critical step involves identifying and selecting key features (or observables) that significantly impact the performance of the models studied. To this end, we employed SHAP to define the importance of various observables within our dataset. As previously noted, the data used to train our QML models was based on pairwise correlations between all spins in the structure.  Specifically, we calculated the expected values of the observables $(\langle\sigma_{i}^{x} \sigma_{j}^{x}\rangle, \langle\sigma_{i}^{y} \sigma_{j}^{y}\rangle, \langle\sigma_{i}^{z} \sigma_{j}^{z}\rangle)$  for $j > i$ and $i = [1, N-1]$, where $N$ represents the number of spins. It is among these features that we apply SHAP to identify the most relevant ones that significantly influence model performance.

Figure \ref{fig:shap1} shows the feature importance ranking for classifying the phases of the ANNNI model using SVC on an 8-site system. { This is known as a SHAP summary plot, where features with SHAP values around zero indicate a low impact on the model's output.}  As we can see, the three most important features identified were $\langle\sigma_1^z, \sigma_4^z\rangle$, $\langle\sigma_1^z, \sigma_3^z\rangle$, and $\langle\sigma_1^z, \sigma_2^z\rangle$, indicating that the spin projection along the $z$-axis ($\sigma^z$) plays a more significant role in phase classification compared to other projections. Physically, this shows the significant role of longitudinal spin correlations in determining the phase structure of the ANNNI model. The strong influence of $\sigma^z$ pairs involving nearest and next-nearest neighbors highlights the importance of these specific interactions in driving phase transitions, particularly in relation to the ordering of spins along the $z$-axis.

\begin{figure}
    \centering
    \includegraphics[width=1\linewidth]{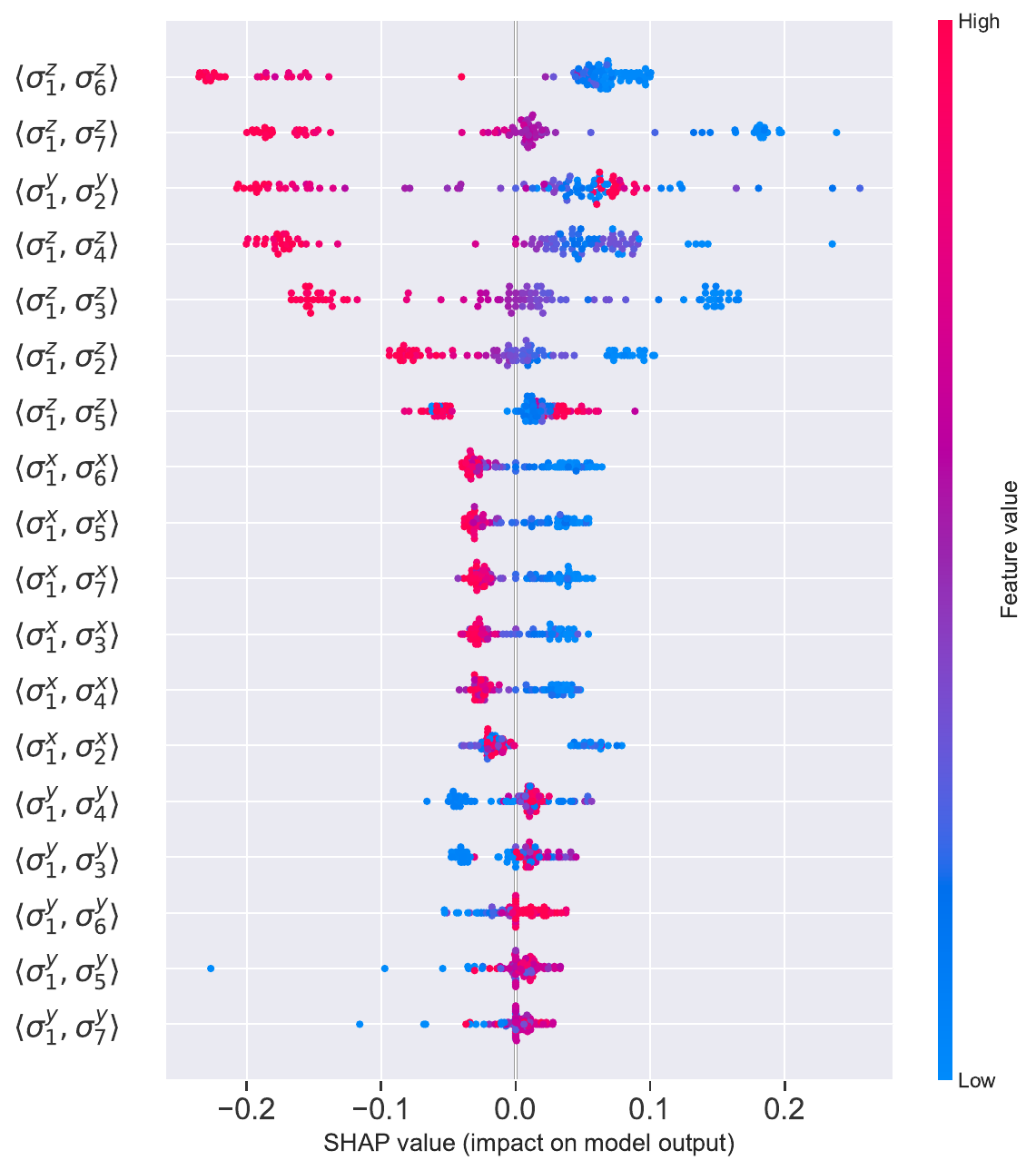}
    \caption{SHAP feature importance ranking for a 12-site ANNNI model. The plot reveals the most significant features for phase classification in this larger system.}
    \label{fig:shap2}
\end{figure}

Figure \ref{fig:shap2} show the results for a 12-spin site model. The most important features identified were $\langle\sigma_1^z, \sigma_6^z\rangle$, $\langle\sigma_1^z, \sigma_7^z\rangle$, and $\langle\sigma_1^y, \sigma_2^y\rangle$. Notably, with the exception of the third feature, all other important features involve $\sigma^z$ correlations. This pattern suggests a consistent dominance of the $\sigma^z$ projection in determining phase behavior. The continued prominence of $\sigma^z$ interactions across different system sizes reinforces their critical role in defining phase transitions within the ANNNI model. Although the contributions of $\sigma^x$ and $\sigma^y$ projections are less pronounced by comparison, they are not negligible, with the $\sigma^y$ projection showing some relevance in one of the cases.

It is important to emphasize that by applying SHAP with a classical SVC, we align the feature selection process with the classical equivalent of the QSVM, potentially giving the QSVM an advantage in terms of feature relevance. This compatibility may result in more meaningful features for QSVM. On the other hand, the VQC lacks a direct classical equivalent,  making it more difficult to accurately determine feature importance within its framework. This discrepancy points to the need for quantum-specific Explainable AI techniques that are better suited to models like the VQC. With key observables now identified through our feature selection process, we can proceed to the next phase of our study: performing classification tasks using these selected features.



\subsection{Classification Results}

In our approach, we implement the QSVM using the fidelity-based quantum kernel with the ZFeatureMap for data encoding. The ZFeatureMap was configured with three repetitions, enabling an effective transformation of classical data into quantum states. To construct the quantum kernel, we employed the FidelityStatevectorKernel \cite{qiskit2024}, which measures the similarity between these quantum states. 
Similarly, we used the VQC with the {ZFeatureMap} for data encoding and the {EfficientSU2} \cite{Kandala_2017} ansatz, utilizing five repetitions of linear entanglement. The circuit parameters were optimized using the SPSA optimizer\cite{Spall_1992} over $100$ iterations, with expectation values estimated through $600$ shots during training. Both models were followed the same training methodology, ensuring a fair comparison, and were evaluated using the same accuracy metric.

The results presented in Table \ref{tab:combined_accuracy_vertical} demonstrate that both QSVM and VQC algorithms were effective in classifying quantum phases, with the accuracy in all cases exceeding 94\%. Notably, QSVM outperformed VQC in the cases studied, particularly in the 8-site system, where it achieved an impressive accuracy of 98.46\%, compared to VQC's 94.81\%. For the 12-site system, the accuracies for QSVM and VQC were slightly lower at 97.73\% and 96.49\%, respectively. This consistent, albeit slight, decrease in accuracy as the system size increases can raise important considerations about the limitations of feature sets in capturing the complex interactions within larger quantum systems. 

\begin{table}[h!]
\centering
\begin{tabular}{ccc}
\toprule
\textbf{System Size} & \textbf{Algorithm} & \textbf{Accuracy} \\
\midrule
8-site  & VQC  & 94.81\% \\
12-site & VQC  & 96.49\% \\
8-site  & QSVM & 98.46\% \\
12-site & QSVM & 97.73\% \\
\bottomrule
\end{tabular}
\caption{Classification performance of QSVM and VQC algorithms on 8-site and 12-site systems, utilizing the top 5 features with the highest impact as determined by the SHAP algorithm.}
\label{tab:combined_accuracy_vertical}
\end{table}

\begin{figure}
    \centering
    \includegraphics[width=\linewidth]{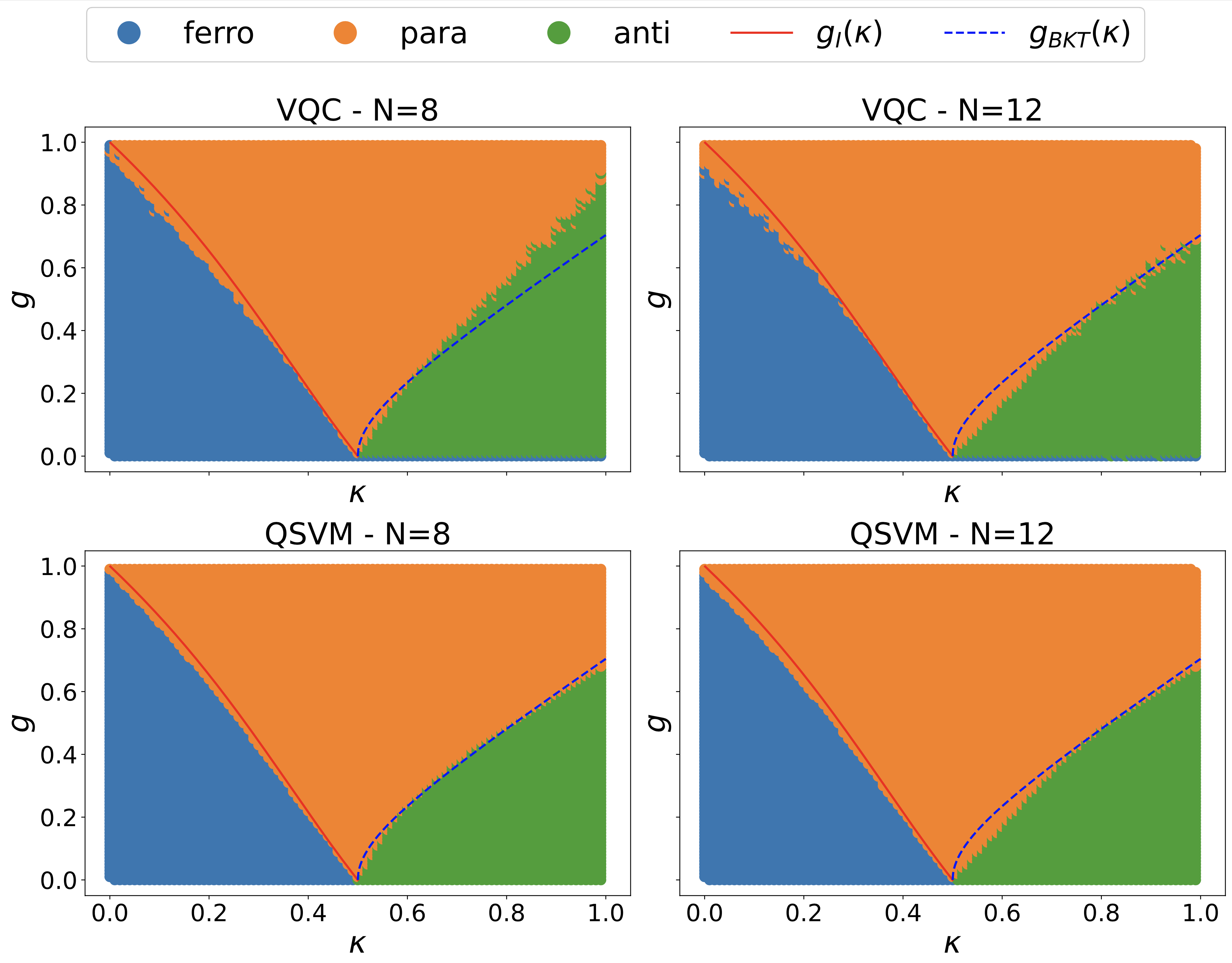}
    \caption{Phase diagram of the ANNNI model, delineated through classification via VQC (Top) and QSVM (Bottom). Presented with 8-site chains (left) and 12-site chains (right). The markers denote phase predictions as per the model, whereas the lines represent the loci of theoretical transitions.}
    \label{fig:diagrams}
\end{figure}

The superior performance of the QSVM algorithm can be attributed to its capacity to more effectively exploit the limited features available, particularly in smaller systems. Figure \ref{fig:diagrams} provides a visual representation of phase classification using the QSVM and VQC for an 8-site and 12-site system with 5 features. As observed, the models accurately classified the phases; however, they encountered difficulties in identifying the phase boundaries, with the 8-site QSVM providing the closest approximation to the actual phase transitions observed in the underlying quantum model. This suggests that the QSVM is more adept at capturing the critical features necessary for phase identification in these specific scenarios. The VQC model, despite its overall competent performance, exhibited some misclassifications, notably along the $ g = 0 $ line for $ k > 0.5 $, where it incorrectly labeled the phase as ferromagnetic rather than antiferromagnetic. This misclassification was consistent across both the 8-site and 12-site systems. Such limitation in phase boundary detection contributed to the slightly lower accuracy observed in the VQC results.

\begin{figure}
    \centering
    \includegraphics[width=\linewidth]{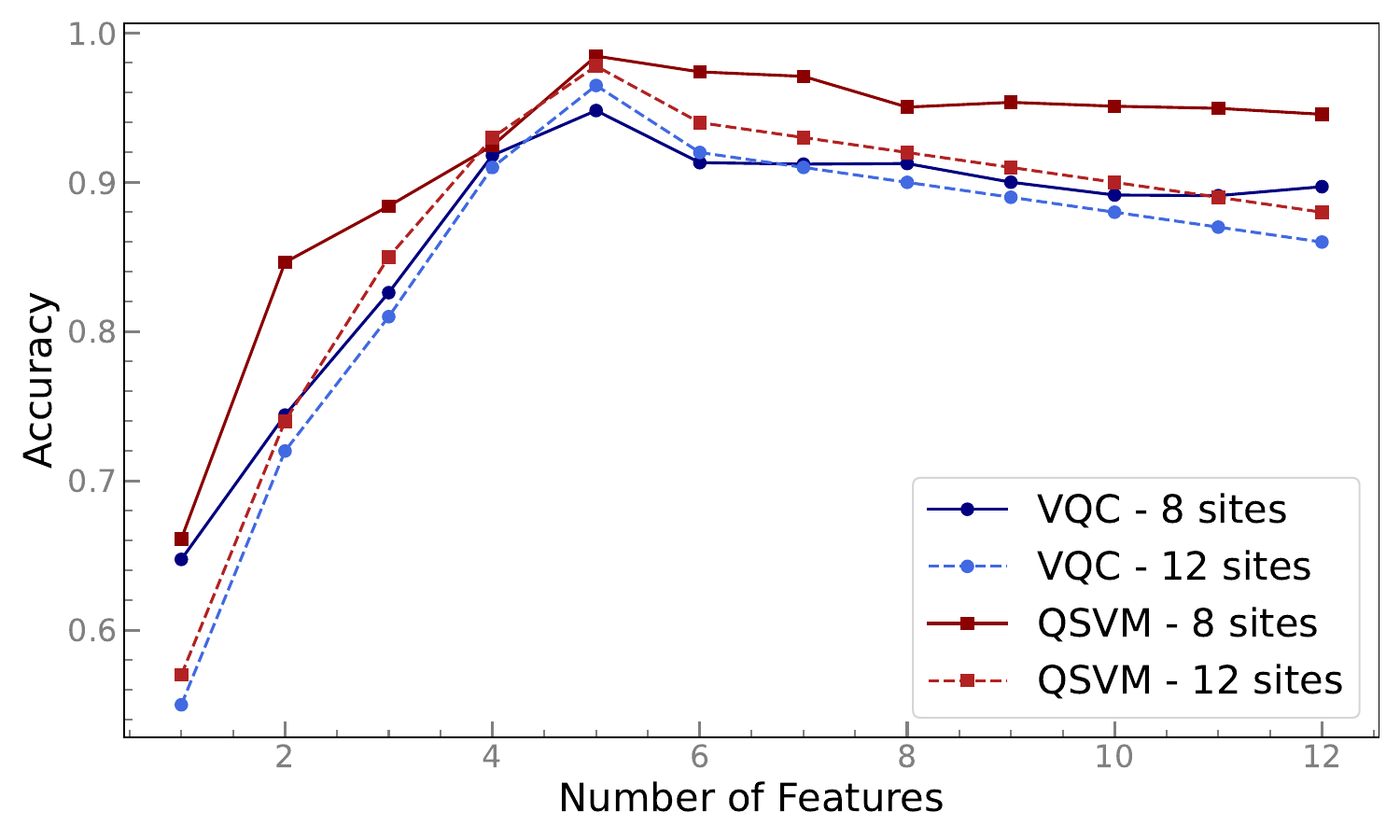}
    \caption{Accuracy as a function of the number of features for QSVM and VQC.}
    \label{fig:accuracy_features}
\end{figure}

Figure \ref{fig:accuracy_features} explores the relationship between the number of selected features and the accuracy of both QSVM and VQC algorithms. The results indicate that QSVM generally outperforms VQC across most feature sets. For both algorithms, the optimal performance was achieved when using 5 features. Beyond this range, the accuracy of both models tended to decline, highlighting the importance of careful feature selection in quantum machine learning tasks. These findings underscore the effectiveness of using SHAP for feature selection in quantum machine learning. SHAP was instrumental in identifying the most relevant features for phase classification, enabling high accuracy even with a reduced feature set. 

In the context of the NISQ (Noisy Intermediate-Scale Quantum) era, where quantum computers are constrained by a limited number of qubits, the ability to achieve high accuracy with a minimal feature set is particularly valuable. The results demonstrate that effective quantum phase classification can be achieved without the need for a large number of features, making QSVM and VQC viable options for phase classification tasks in current and near-future quantum computing environments.

\section{Conclusion}
\label{conclusion}
This study shows that leveraging feature selection via SHAP facilitates the achievement of high accuracy in quantum phase classification utilizing QML algorithms, even with a minimal feature set. The efficacy of SHAP in identifying the most pertinent features enabled both QSVM and VQC algorithms to achieve precise classification of quantum phases. This capability is of particular merit in the NISQ era, characterized by the constraints imposed by the limited availability of qubits within quantum computing resources.

The capacity to sustain elevated classification accuracy with a diminished feature set underscores the prospective efficacy of QSVM and VQC as formidable instruments for quantum phase classification in environments constrained by resources. This methodology not only increases the computational efficiency of quantum processes but also facilitates the progression of QML algorithms to more intricate quantum systems. 
Subsequent research endeavors should persist in the refinement of feature selection techniques and in exploring their extensive applications within quantum machine learning, thereby ensuring that these algorithms are proficiently utilized within the progressively advancing domain of quantum computing.

\vspace{1cm}
\begin{acknowledgments}
 G.S.F acknowledge support from Conselho Nacional de Desenvolvimento Cient\'{i}fico e Tecnol\'{o}gico (CNPq), project number 2023/9445. F. M. acknowledges the support of Coordena\c{c}{\~a}o de Aperfei\c{c}oamento de Pessoal de N{\'i}vel Superior (CAPES), project number 88887.607339/2021-00. P.M.P and F.F.F acknowledge support from Funda\c{c}{\~a}o de Amparo {\`a} Pesquisa do Estado de S{\~a}o Paulo (FAPESP), project numbers 2023/12110-7 and 2023/04987-6 respectively.  F.F.F. acknowledges support from ONR, Project No. N62909-24-1-2012.
\end{acknowledgments}

\bibliography{bibl}

\end{document}